\documentclass[runningheads]{llncs}
\usepackage{perpage} 
\MakePerPage{footnote} 
\usepackage{graphicx}
\usepackage{amsmath}
\usepackage{multirow}
\usepackage{booktabs}
\usepackage{hyperref}
\usepackage{xcolor}
\usepackage{enumitem}
\usepackage{tabularx}
\usepackage{array}
\usepackage{subcaption}
\usepackage{physics}
\usepackage{amssymb}

\newcolumntype{C}{>{\centering\arraybackslash}X}

\begin{document}

\title{DREQ: Document Re-Ranking Using Entity-based Query Understanding}

\titlerunning{DREQ: Document Re-Ranking Using Entity-based Query Understanding}

\author{Shubham Chatterjee\inst{1}\orcidID{0000-0002-6729-1346} 
\and
Iain Mackie\inst{2}
\and \\
Jeff Dalton\inst{3}\orcidID{0000-0003-2422-8651}}
\authorrunning{S. Chatterjee et al.}
\institute{University of Edinburgh, UK
\email{shubham.chatterjee@ed.ac.uk}
\and
University of Glasgow, UK
\email{i.mackie.1@research.gla.ac.uk}
\and
University of Edinburgh, UK
\email{jeff.dalton@ed.ac.uk}
}
\maketitle              

\begin{abstract}

While entity-oriented neural IR models have advanced significantly, they often overlook a key nuance: the varying degrees of influence individual entities within a document have on its overall relevance. Addressing this gap, we present \texttt{DREQ}, an entity-oriented dense document re-ranking model. Uniquely, we emphasize the query-relevant entities within a document's representation while simultaneously attenuating the less relevant ones, thus obtaining a query-specific entity-centric document representation. 
We then combine this entity-centric document representation with the  text-centric representation of the document to obtain a ``hybrid'' representation of the document. We learn a relevance score for the document using this hybrid representation. Using four large-scale benchmarks, we show that \texttt{DREQ} outperforms state-of-the-art neural and non-neural re-ranking methods, highlighting the effectiveness of our entity-oriented representation approach. 
\end{abstract}
\section{Introduction}
\label{sec:Introduction}

During the last decade, the emergence of large-scale Knowledge Graphs (KGs) has motivated the development of entity-oriented search systems. 
Entities, with their rich semantic information, help bridge the gap between unstructured text and structured knowledge.
Prior research \cite{ensan2017document,xiong2015esdrank,xiong2017word,liu2015latent,raviv2016document} underscores the significance of entities in feature-based retrieval systems. Their effectiveness within neural IR models has also been demonstrated, with the Entity-Duet Ranking Model (EDRM) from Liu et al. \cite{liu-etal-2018-entity} standing out as a pioneering effort. This model synergizes the word-entity duet framework \cite{xiong2017word} with the strengths of neural networks and KGs. More recently, Tran and Yates \cite{tran2022dense} introduced an approach that clusters entities within documents, offering multiple ``views'' or perspectives to enhance the understanding of various document facets.

Yet, amidst these advancements, 
current models often overlook a crucial aspect: 
not all entities within a document contribute equally to its relevance. 
For instance, given the query ``Black Bear Attacks'' and the document 
in Figure \ref{fig:example-intro}, while the entity \textit{hand axe} may have peripheral relevance to the given query due to its defensive use during animal encounters, the entity \textit{National Guard} is likely non-relevant. Conversely, 
the entity \textit{Parks Highway}, known for bear sightings, may hold greater significance. 
%
In the ``retrieve-then-rerank'' paradigm, a common approach in neural IR, initial retrieval fetches a broad set of candidate documents. For the subsequent re-ranking phase, though, the differential relevance of individual entities becomes particularly critical as the model must sift through the candidates and rank them with high precision. 
Furthermore, prior work \cite{gabrilovich2009wikipedia,xiong2017word,liu-etal-2018-entity,tran2022dense} often produce query-agnostic document representations which fail to resonate with the specific nuances and requirements of the query. 
Our proposition emphasizes a more refined approach: to enhance re-ranking accuracy, entities should be weighted based on their query relevance, ensuring that the document's representation is both influenced by the most relevant entities and tailored to the query's nuances.

Against this backdrop, in this work\footnote{\textbf{Code and data: \url{https://github.com/shubham526/ECIR2024-DREQ}.}}, we introduce \textit{\textbf{D}ocument \textbf{R}e-ranking using \textbf{E}ntity-based \textbf{Q}uery Understanding} (\texttt{DREQ}), an entity-oriented neural re-ranking model that extends the conventional ``retrieve-then-rerank'' paradigm by introducing an innovative intermediate step. Given a query $Q$ and a candidate set of documents $\mathcal{D}$ retrieved using an initial retrieval method (e.g., BM25), we want to \textit{re-rank} these candidates to order them by their relevance to the query $Q$.  
While prior approaches predominantly utilize entities to gain a fine-grained understanding of documents, our method uniquely perceives entities within candidate documents as overarching concepts essential for a comprehensive understanding of the query. 
We identify and prioritize entities that align closely with the query. To this end, we emphasize the embeddings of the relevant entities and concurrently attenuate the less relevant ones within the document's representation, thus obtaining a refined, \textit{query-specific} and \textit{entity-centric} document representation. Recognizing that the raw text of a document captures the document's overarching semantic context, we also derive a broader text-centric representation of the document, complementing the focused entity-centric perspective. We meld the entity-centric and text-centric representations of the document, obtaining a ``hybrid'' document representation that imbibes insights about what the query specifically seeks. Using this hybrid representation, we learn a fine-grained ``interaction vector'' that imbibes the differences, commonalities, and subtle relationships between the query and document. We then use this vector to learn a relevance score for the document. 
We term our methodology ``retrieve-harness-rerank''\footnote{Term coined by Dr. Laura Dietz at the SIGIR 2023 tutorial  (\url{https://github.com/laura-dietz/neurosymbolic-representations-for-IR}).}, emphasizing our process of harnessing the information from entities present in candidate documents to re-rank the candidates.

\paragraph{\textbf{Contributions.}} We make the the following contributions through this work:

\begin{enumerate}
    \item We introduce \texttt{DREQ}, an entity-oriented re-ranking model that enriches a document's representation with \textit{query-specific} entity knowledge for nuanced relevance matching. We achieve new state-of-the-art results on four major document ranking test collections. 

    \item We introduce a hybrid representation learning mechanism, blending entity-centric and text-centric representations. This hybrid representation captures a document's broad context and query-specific relevance,  enhancing the precision of re-ranking. 

    \item We show that the effectiveness of re-ranking is significantly enhanced by meticulously selecting and assigning appropriate weights to entities within a document.
\end{enumerate}

\section{Related Work}
\label{sec:Related Work}


\textbf{Entity-Oriented Search.} Initial attempts at entity-oriented search primarily used entities for query expansion, one notable example being Entity-Query Feature Expansion \cite{dalton2014entity} (EQFE) model which used entity links within documents for query expansion. 
Entities later became a latent layer \cite{gabrilovich2009wikipedia,liu2015latent,xiong2015esdrank} in document and query representations, forming a high-dimensional entity space that improved retrieval by revealing hidden semantics.
Research progressed to treat entities as explicit elements in retrieval models, coexisting with term-based approaches. Methods like entity-based language models \cite{raviv2016document} and semantically-driven models \cite{ensan2017document} rank documents by their semantic relation to the query.
A prominent line of research \cite{xiong2017explicit,xiong2017word,xiong2017jointsem,xiong2015esdrank,xiong2018towards} introduced a dual-layered approach, combining a ``bag-of-entities'' with the traditional ``bag-of-terms'' to improve document retrieval.

\textbf{Neural IR.} Recently, deep learning has transformed text ranking, removing the need for handcrafted features and fostering semantic matching. Before BERT \cite{devlin2018bert}, models either created vector representations \cite{huang2013learning,shen2015entity,mitra2019an,nalisnick2016improving} for queries and documents  or built similarity matrices to capture term interactions \cite{guo2016drmm,xiong2017end,hui-etal-2017-pacrr,hui2018copacrr,dai2018cknrm}. The advent of BERT  and its derivatives \cite{liu2019roberta,clark2020electra,he2020deberta,zhang-etal-2019-ernie,jiang2020convbert} in a retrieve-then-rerank \cite{nogueira2019passage,nogueira2019multi} framework ushered in models like Birch \cite{akkalyoncu-yilmaz-etal-2019-cross} and BERT-MaxP \cite{dai2019deeper}, emphasizing sentence or passage relevance. Subsequent models, such as CEDR \cite{macavaney2019ceder} and PARADE \cite{li2020parade}, utilized BERT's contextual embeddings, with the latter aggregating passage representations. Concurrently, ERNIE \cite{zhang-etal-2019-ernie} enhanced BERT with knowledge graphs and vast text corpora.

Meanwhile, bi-encoders emerged as a foundational mechanism in text ranking, employing distinct encoders to derive query and document vectors. For example, DPR \cite{karpukhin-etal-2020-dense} employs BERT's \texttt{[CLS]} token for both, gauging similarity through their inner product. The choice of negative examples is crucial: DPR favors BM25-retrieved passages while ANCE \cite{xiong2020ance} leans on ANN techniques. Addressing bi-encoders' challenges with term-level interactions, research has delved into multi-vector text representation, exemplified by ColBERT's \cite{khattab2020colbert} ``late interaction'' using token-level embeddings. 

The evolution of pre-trained sequence-to-sequence models like T5 \cite{raffel2019exploring} has led to models such as MonoT5 \cite{nogueira2019document} and RankT5 \cite{zhuang2023rankt5}: The former assesses document relevance via probability assignment to ``true'' labels during decoding whereas the latter provides direct ranking scores for query-document pairs.

\textbf{Query/Document Expansion.} 
Recent efforts in this direction have been driven by embeddings. Solutions such as DeepCT \cite{dai2020context} and docT5query \cite{nogueira2019document} use transformers to enrich traditional PRF models by highlighting significant terms in documents. Simultaneously, methods like Neural PRF \cite{li-etal-2018-nprf} and BERT-QE \cite{zheng-etal-2020-bert} have adopted neural ranking models to measure document similarity with feedback documents. CEQE \cite{naseri2021ceqe} has further enhanced this approach by using BERT to create contextualized representations and select expansion terms based on similarity in the embedding space. While ColBERT-PRF \cite{xiao2023colbert-prf} directly utilizes BERT embeddings for retrieval without further training,  ANCE-PRF \cite{hongchien2021ance-prf} requires additional retraining of the query encoder using PRF information

\textbf{Entity Ranking.} 
Early methods include using MRFs to deal with the joint distribution of entity terms from semi-structured data ~\cite{metzler2005mrf,zhiltsov2015fielded,nikolaev2016paramterized,hasibi2016exploiting,raviv2012aranking}, utilizing types  ~\cite{kaptein2010entity,balog2011query,garigliotti2017on} and relations ~\cite{tonon2012combining,ciglan2012the} from a KG, and using Learning-To-Rank methods ~\cite{schuhmacher2015ranking,graus2016dynamic,dietz2019ent,chatterjee2021fine} to rank entities using a diverse set of features.
Recent advancements have been significantly driven by neural models using advanced techniques such as autoregressive entity ranking \cite{cao2021autoregressive}, integrating BERT-based entity rankings with other features \cite{chatterjee2022berter}, and enhancing BERT with Wikipedia2Vec \cite{yamada-etal-2020-wikipedia2vec} embeddings \cite{gerritse2022embert}.
Concurrently, the emergence of graph embedding-based models ~\cite{gerritse2020graph,nikolaev2020joint} 
has enriched entity ranking approaches by utilizing joint embedding of entities and words within the same vector space. 

\section{Approach: DREQ}
\label{sec:DREQ: Approach}

\begin{figure}[t]
    \centering
    \includegraphics[width=1\linewidth]{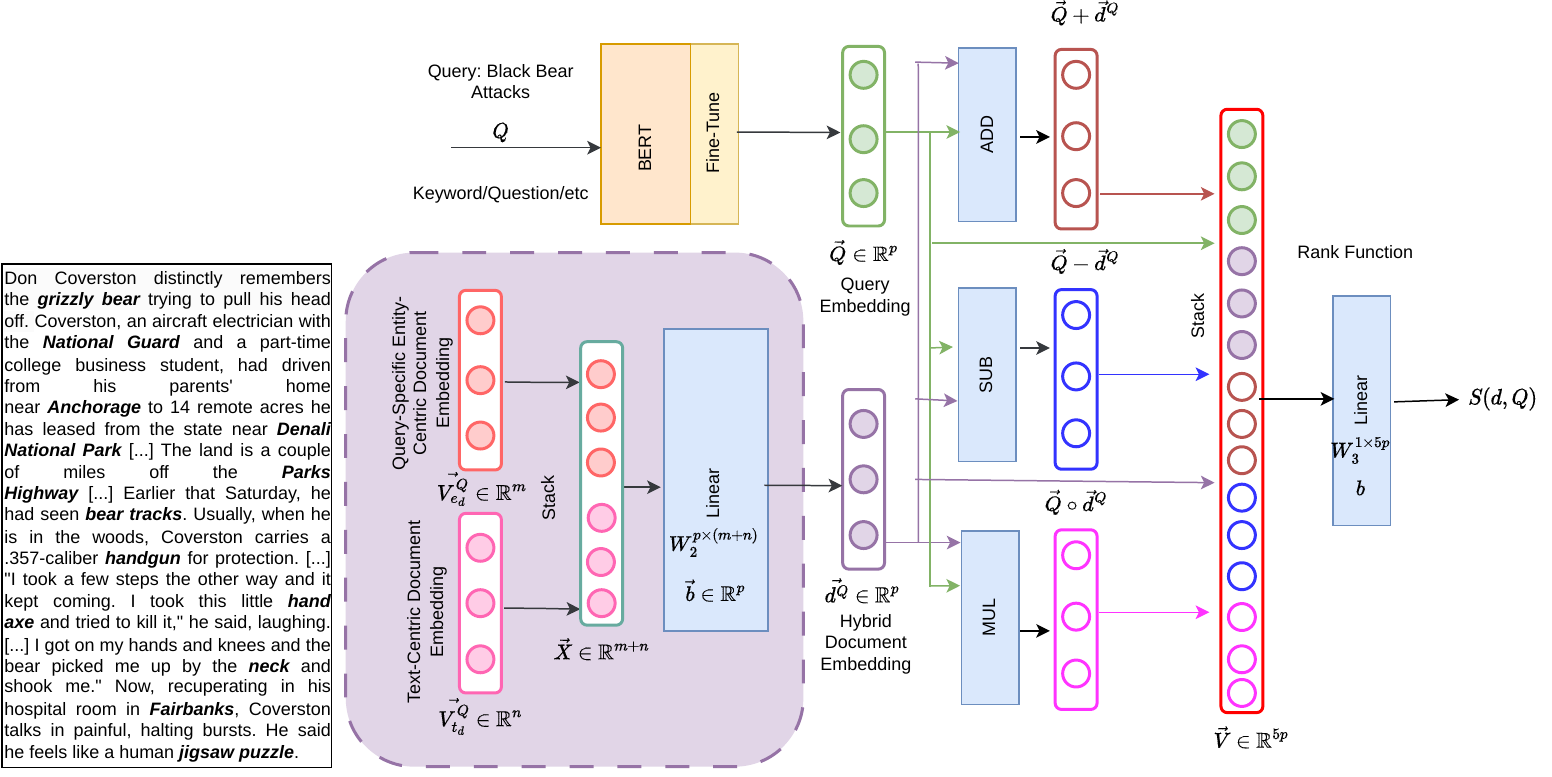}
    \caption{Our proposed system \texttt{DREQ} uses a hybrid document embedding learnt using (1) the query-specific entity-centric embedding, and (2) text embedding of the document to learn the document score. 
    }
    \label{fig:example-intro}
\end{figure}

\paragraph{\textbf{Overarching Idea.}} 
Our work is anchored in the intuition that documents contain interconnected entities which provide a distilled understanding of the document's main content, and serve as broad concepts important for understanding the query.  For example, in Figure \ref{fig:example-intro}, given the query ``Black Bear Attacks'', the presence of specific geographical entities such as \textit{Parks Highway} and \textit{Fairbanks} narrows down the context to the Alaskan wilderness, a region known for its bear population. 
Additionally, 
the entity \textit{Hand axe} sheds light on human defensive measures during such encounters.
Together, these entities reinforce the document's fit for narratives around bear encounters.  
Even though the central narrative revolves around a grizzly, the web of interconnected entities creates an ecosystem of information that can be invaluable in assessing relevance to the broader theme of bear attacks.

\paragraph{\textbf{Entity Ranking}.} Nonetheless, the importance of these entities for determining the document's relevance is query-dependent. For example, in the context of ``Black Bear Attacks'', the entity \textit{Parks Highway} is highly important as it specifies a geographical context known for its black bear population.
However, 
for a broader query like ``Human-animal conflict in North America'', the entity is less significant as the theme encompasses a vast range of conflicts, animals, and locations, making the specific location of \textit{Parks Highway} just one of many possible locales of interest and less indicative of the document's relevance.

To address this issue, we pool all entities from all candidate documents $d \in \mathcal{D}$ to obtain a candidate set of entities $\mathcal{E}$ for the query $Q$. Following previous work \cite{nguyen2016marco,dietz2017trec,dietz2018trec}, we transfer the relevance labels from documents to entities in the document based on the assumption that a relevant document contains relevant entities. Using this entity ground truth, we train a separate entity ranking model to rank entities $e \in \mathcal{E}$. Specifically, we follow previous work ~\cite{manotumruksa2020crossbert,xiong2016esr,liu-etal-2018-entity,yamada-etal-2018-representation,chatterjee2022berter} and leverage the Knowledge Base (DBpedia \cite{lehmann2015dbpedia}) description $t_e$ of entity $e$ to learn an embedding $\vb{e} \in \mathbb{R}^k$ of the entity using BERT. 
The input to BERT is a sequence of query tokens $\tau^q \in Q$ and description tokens $\tau^e \in t_e$, separated by the special token \texttt{[SEP]}, and preceded by the special token \texttt{[CLS]}. 
We use the $k$-dimensional embedding of the \texttt{[CLS]} token from the last hidden layer of BERT as the embedding $\vb{e}$ of an entity $e \in \mathcal{E}$. To derive a rank score for this entity in relation to the query $Q$, we learn a linear 
projection $\mathbb{R}^k \rightarrow \mathbb{R}$ using the embedding $\vb{e}$ and a weight matrix $W_1^{1 \times k}$. This scoring function $\text{S}(e,Q)$ is formulated as: $\text{S}(e,Q) = W_1 \cdot \vb{e} + b$, 
where $b$ is a scalar bias term.


\paragraph{\textbf{Document Representation.}} 
Entities such as \textit{Parks Highway}, \textit{Neck}, and \textit{Hand axe} in Figure \ref{fig:example-intro} collectively suggest a bear encounter in Alaska. When related to the query ``Black Bear Attacks'', the combined narrative from the entities implies the document is probably relevant as it aligns with the query's essence.
%
%
Based on this idea, and acknowledging that the significance of these entities for determining a document's relevance varies with the query, we learn a \textbf{\textit{query-specific entity-centric}} representation $\vb{V}_{e_d}^Q \in \mathbb{R}^m$ of a candidate document $d \in \mathcal{D}$. Specifically, we first represent each entity within a document via its embedding from Wikipedia2Vec \cite{yamada-etal-2020-wikipedia2vec} due to its ability to capture the relationships and deeper contexts between entities. 
The document representation $\vb{V}_{e_d}^Q$ is a weighted sum 
of embeddings of entities in the document: 
\begin{equation}
\label{eq:entity-centric-document-emb}
    \vb{V}_{e_d}^Q = \sum_{e \in d} w_e \cdot \vb{e}
\end{equation}
where the weight $w_e$ for each entity is query-specific, determined by the rank score of the entity: 
\begin{equation*}
    w_e = \frac{S(e,Q)}{\sum_{e' \in d} S(e',Q)}
\end{equation*}
Note that $S(e,Q)$ are normalized using the \texttt{softmax} function. Consequently, the weights $w_e$ can be interpreted as the \textit{probability} of the entity's relevance to the query. Intuitively, using these weights would prioritize query-relevant entities while concurrently down-weighting those of lesser relevance within the document. 

While encapsulating key entities helps capture a document's query-specific semantic essence, it's crucial to also consider the broader textual context to capture nuances missed by solely focusing on entities.
%
%
Hence, we also obtain a text-centric representation $\vb{V}_{t_d}^Q \in \mathbb{R}^n$ of a candidate document $d \in \mathcal{D}$ using BERT. 
Following previous work \cite{macavaney2019ceder,li2020parade}, we first segment the document into passages using a sliding window of $M$ sentences with a stride of $S$ sentences over the document. We represent each passage in the document via the embedding of the \texttt{[CLS]} token from BERT. To obtain the document representation $\vb{V}_{t_d}^Q$, we average the passage embeddings. 
%
We combine the text and entity-centric document representations 
to learn a single \textbf{\textit{hybrid}} representation $\vb{d}^Q \in \mathbb{R}^p$ via a linear 
projection $\mathbb{R}^{m+n} \rightarrow \mathbb{R}^p$ 
as follows: 
\begin{equation*}
    \vb{d}^Q = W_2 \cdot [\vb{V}_{t_d}^Q; \vb{V}_{e_d}^Q] + \vb{b}
\end{equation*}
where $\vb{X} \in \mathbb{R}^{m+n}$ 
is the concatenated embeddings $[\vb{V}_{t_d}^Q; \vb{V}_{e_d}^Q]$, $W_2^{p \times (m + n)}$ is a weight matrix, and $\vb{b} \in \mathbb{R}^p$ is a bias vector. 
Our intuition is that this hybrid embedding $\vb{d}^Q$ of a candidate document $d \in \mathcal{D}$ encapsulates both the granular insights from the entities and the overarching narrative provided by the document's text, thereby ensuring a more accurate search.

\paragraph{\textbf{Document Ranking.} }
To learn the document scoring function $\text{S}(d,Q)$, we first learn several fine-grained interactions between the query embedding $\vb{Q} \in \mathbb{R}^p$ (obtained via the embedding of the \texttt{[CLS]} token from BERT) and the hybrid document embedding $\vb{d}^Q$ as follows: An additive interaction $\vb{V_{add}^{d,Q}} = \vb{Q} + \vb{d^Q}$, subtractive interaction $\vb{V_{sub}^{d,Q}} = \vb{Q} - \vb{d^Q}$, and multiplicative (Hadamard product) interaction $\vb{V_{mul}^{d,Q}} = \vb{Q} \circ \vb{d^Q}$.
We then learn the scoring function $\text{S}(d,Q)$ through a linear projection $\mathbb{R}^{5p} \rightarrow \mathbb{R}$ as follows: 
\begin{equation*}
    \text{S}(d,Q) = W_3 \cdot \vb{V} + b
\end{equation*}
where $W_3^{1 \times 5p}$ is a weight matrix, $\vb{V} \in \mathbb{R}^{5p}$ is a vector representing the concatenated embeddings $[\vb{Q};\vb{d^Q};\vb{V_{add}^{d,Q}};\vb{V_{sub}^{d,Q}};\vb{V_{mul}^{d,Q}}]$, and $b$ is a scalar bias.
We hypothesize that by subtracting, adding, and multiplying the query and hybrid document embeddings, we can explore different aspects of how the query relates to the document. Subtraction might show the differences between them, which can help point out any areas that don't match. On the other hand, addition might reveal what they have in common, highlighting overlapping themes. Multiplying the embeddings goes a step further and might reveal subtle relationships between specific parts of the query and document. By merging these interactions with the original query and document embeddings, our intuition is that the model would garner a rich, composite insight into the document's depth and breadth. 

\paragraph{\textbf{End-To-End Training.}} We train both our entity and document ranking models using the binary cross-entropy loss below:
\begin{equation*}
    \mathcal{L} = - \frac{1}{N} \sum_{i=1}^N \biggl(y_i \cdot \text{log} (p(\hat{y}_i)) + (1-y_i) \cdot (1-\text{log}(p(\hat{y}_i)))\biggr)
\end{equation*}
where $y_i$ is the label and $p(\hat{y}_i)$ is the predicted probability of the entity/document being relevant. 
This is analogous to pointwise learning-to-rank where each of $N$ query-entity/query-document pairs is independently classified into relevant or non-relevant.
The entire model is optimized end-to-end using back-propagation. The weight matrices, namely $W_1$ for the entity ranking model and $(W_2, W_3)$ for the document ranking model, are learnt concurrently with the fine-tuning of the initial embeddings. Moreover, the Wikipedia2Vec embeddings undergo end-to-end fine-tuning specifically within the document ranking model, ensuring that these entity embeddings are tailored to enhance document ranking effectiveness.

\section{Experimental Methodology}
\label{sec:Experimental Methodology}

\subsection{Datasets}
\label{subsec:Datasets}

\paragraph{\textbf{CODEC.}} CODEC \cite{mackie2022codec} is a benchmark specifically designed for complex research topics in social sciences. The benchmark includes 42 topics, and new entity linked corpus with 729,824 documents with focused content across finance, history, and politics. The corpus contains an average of 159 entities per document.
 It provides expert judgments on 6,186 documents derived from diverse automatic and manual runs.

\paragraph{\textbf{TREC Robust 2004.}} The TREC Robust 2004 track focuses on poorly performing topics. The track provides 250 topics with short ``titles'' and longer ``descriptions''; we report results on both title and description queries. The collection consists of 528,024 documents (containing an average of 116 entities per document) taken from TREC disks 4 and 5 excluding the Congressional Record. The track provides 311,409 graded relevance judgments for evaluation.

\paragraph{\textbf{TREC News 2021.}} The TREC News track focuses on search tasks within the news domain. We focus on the background linking task, which involves retrieving news articles that provide relevant context or background information for a given news story. There are 51 topics, each with a title, description, and narrative; in this work, we use all three fields for query formulation. The track uses the TREC Washington Post (v4) collection, encompassing 728,626 documents containing 131 entities per document on average. The track provides 12,908 graded relevance assessments for evaluation.

\paragraph{\textbf{TREC Core 2018.}} The TREC Core track offers 50 topics, each consisting of titles, descriptions, and narratives. For this work, we utilize all three components. The track uses the TREC Washington Post (v2) collection, encompassing 595,037 news articles and blog posts and containing 123 entities per document on average. 26,233 graded relevance judgements are available.

\subsection{Evaluation Paradigm}
\label{subsec:Evaluation Paradigm}

\paragraph{\textbf{Candidate Ranking.}} We retrieve a candidate set of 1000 documents per query using BM25+RM3 (Pyserini default). The Recall@1000 of this candidate set for each dataset is as follows: (1) CODEC: 0.82, (2) Robust04 (title): 0.77, (3) Robust04 (desc): 0.75, (4) News 2021: 0.94, (5) Core 2018: 0.75. Other metrics shown in  Tables \ref{tab:res-tab-1} and \ref{tab:res-tab-2}


\paragraph{\textbf{Evaluation Metrics.}} (1) Precision at $k=20$, (2) Normalized Discounted Cumulative Gain (nDCG) at $k=20$, and (3) Mean Average Precision (MAP). 
We conduct significance testing using paired-t-tests. 



\paragraph{\textbf{Entity Linking.}} Our work relies on an entity linked corpus. While any entity linking system may be used, in this work, we use WAT \cite{piccinno2014wat}. 

\paragraph{\textbf{Train and Test Data.}}  As positive examples during training, we use documents that are assessed as relevant in the ground truth provided with the dataset. Following the standard \cite{karpukhin-etal-2020-dense}, for negative examples, we use documents from the candidate ranking (BM25+RM3) which are either explicitly annotated as negative or not present in the ground truth. We balance the training data by keeping the number of negative examples the same as the number of positive examples. These examples are then divided into 5-folds for cross-validation. We create these folds at the query level.




\paragraph{\textbf{Baselines.}}
We compare our proposed re-ranking approach \texttt{DREQ} to the following supervised state-of-the-art neural re-rankers: (1) \textbf{RoBERTa \cite{liu2019roberta}}, (2) \textbf{DeBERTa \cite{he2020deberta}}, (3) \textbf{ELECTRA \cite{clark2020electra}}, (4) \textbf{ConvBERT \cite{jiang2020convbert}}, (5) \textbf{RankT5 \cite{zhuang2023rankt5}}, (6) \textbf{KNRM ~\cite{xiong2017end}}, (7) \textbf{ERNIE \cite{zhang-etal-2019-ernie}}, (8) \textbf{EDRM \cite{liu-etal-2018-entity}}. Furthermore, we also include an unsupervised entity-based baseline (9) \textbf{MaxSimCos} which scores documents using the maximum cosine similarity between every pair of query and document entity embedding. 
%
On TREC Robust 2004, we include the following additional (full-retrieval) baselines: (1) CEDR \cite{macavaney2019ceder}, (2) EQFE \cite{dalton2014entity}, (3) BERT-MaxP\cite{dai2019deeper}, and (4) PARADE \cite{li2020parade}. 
All baselines are fine-tuned on the target datasets via 5-fold cross-validation using the binary cross-entropy loss.

\subsection{Implementation Details}
\label{subsec:Implementation Details}

We use the \texttt{bert-base-uncased} model from HuggingFace to obtain the initial query and document/entity representations and fine-tune our model using the \texttt{CrossEntropyLoss} function from PyTorch. We use the Adam \cite{kingma2014adam} optimizer with a learning rate of  $10^{-5}$ and batch size of 20. BERT layers were not frozen during fine-tuning. For document segmentation, we apply a 10-sentence sliding window with a 5-sentence stride using spaCy (v6.3.1). Embeddings (document text, and entity) and entity links are are cached off-line for lookup during inference. 
\section{Results and Discussions}
\label{sec:Results and Discussions}


\subsection{Effectiveness of \texttt{DREQ}}
\label{subsec:Effectiveness of DREQ}

\begin{table}[t]
\centering
\caption{Overall results on TREC Robust 2004. Best results in bold. $\blacktriangle$ denotes significant improvement and $\blacktriangledown$ denotes significant deterioration compared to $\star$. Paired-t-test at $p<0.05$. \textbf{nDCG and Precision measures at cut-off rank 20}. Unavailable results denoted by ``--''. Baselines denotes by $^\dagger$ are re-ranking the BM25+RM3 candidate set at the top.
}
\label{tab:res-tab-1}
\scalebox{0.75}{
\begin{tabularx}{\textwidth}{|l|l|CCC|CCC|}
\cline{1-8}
            & & \multicolumn{3}{c|}{\textbf{TREC Robust 2004 (Title)}} & \multicolumn{3}{c|}{\textbf{TREC Robust 2004 (Desc)}} \\ \cline{1-8}
            & & \textbf{MAP}         & \textbf{nDCG}        & \textbf{Prec} & \textbf{MAP}         & \textbf{nDCG}        & \textbf{Prec}        \\ \cline{1-8}
& BM25+RM3     
& 0.29$^\star$           & 0.44$^\star$           & 0.38$^\star$ & 0.28$^\star$	& 0.42$^\star$	& 0.37$^\star$       \\ \cline{1-8}
\multirow{9}{*}{Non-entity} & RankT5$^\dagger$  
& 0.30           & 0.50$^\blacktriangle$            & 0.43$^\blacktriangle$ & 0.33$^\blacktriangle$	& 0.54$^\blacktriangle$	& 0.46$^\blacktriangle$      \\ 
& RoBERTa$^\dagger$     
& 0.29           & 0.47$^\blacktriangle$           & 0.41$^\blacktriangle$ &0.33$^\blacktriangle$	&0.54$^\blacktriangle$	&0.46$^\blacktriangle$       \\
& DeBERTa$^\dagger$      
& 0.29           & 0.49$^\blacktriangle$           & 0.42$^\blacktriangle$   &0.34$^\blacktriangle$	&0.55$^\blacktriangle$	&0.47$^\blacktriangle$     \\
& ELECTRA$^\dagger$     
& 0.27$^\blacktriangledown$           & 0.45           & 0.39  &0.29	&0.49$^\blacktriangle$	&0.41$^\blacktriangle$      \\ 
& ConvBERT$^\dagger$     
& 0.32$^\blacktriangle$           & 0.52$^\blacktriangle$           & 0.45$^\blacktriangle$ &0.35$^\blacktriangle$	&0.57$^\blacktriangle$	&0.48$^\blacktriangle$       \\
& KNRM$^\dagger$         
& 0.11$^\blacktriangledown$           & 0.18$^\blacktriangledown$           & 0.16$^\blacktriangledown$  &0.08$^\blacktriangledown$	&0.14$^\blacktriangledown$	&0.12$^\blacktriangledown$  \\
& CEDR    
& 0.37$^\blacktriangle$           & 0.55$^\blacktriangle$           & 0.48$^\blacktriangle$ &0.40$^\blacktriangle$	&0.60$^\blacktriangle$	&0.52$^\blacktriangle$       \\
& PARADE    
& 0.30           & 0.53$^\blacktriangle$           & 0.45$^\blacktriangle$ &0.30$^\blacktriangle$	&0.56$^\blacktriangle$	&0.47$^\blacktriangle$       \\
& BERT-MaxP    
& 0.32$^\blacktriangle$           & 0.48$^\blacktriangle$           & 0.42$^\blacktriangle$ &0.31$^\blacktriangle$	&0.49$^\blacktriangle$	&0.22$^\blacktriangledown$     
\\ \cline{1-8}
\multirow{4}{*}{Entity-based} & ERNIE$^\dagger$        
& 0.29           & 0.48$^\blacktriangle$           & 0.41$^\blacktriangle$ &0.33$^\blacktriangle$	&0.54$^\blacktriangle$	&0.45$^\blacktriangle$       \\ 
& EDRM$^\dagger$        
& 0.07$^\blacktriangledown$           & 0.10$^\blacktriangledown$            & 0.09$^\blacktriangledown$ &0.05$^\blacktriangledown$	&0.07$^\blacktriangledown$	&0.07$^\blacktriangledown$       \\ 
& EQFE        
& 0.33$^\blacktriangledown$           & 0.42$^\blacktriangledown$            & 0.38$^\blacktriangledown$ & --	& --	& --       \\ 
& MaxSimCos$^\dagger$   
& 0.17$^\blacktriangledown$           & 0.26$^\blacktriangledown$           & 0.24$^\blacktriangledown$  &0.13$^\blacktriangledown$	&0.20$^\blacktriangledown$	&0.18$^\blacktriangledown$      \\ \cline{1-8}
& TREC Best    
& 0.33$^\blacktriangle$           & --           & -- &0.33$^\blacktriangle$	& --	& --       \\ \cline{1-8}
& \textbf{DREQ}        
& $\mathbf{0.57^\blacktriangle}$            & $\mathbf{0.75}^\blacktriangle$           & $\mathbf{0.73}^\blacktriangle$ & $\mathbf{0.55}^\blacktriangle$            & $\mathbf{0.78}^\blacktriangle$           & $\mathbf{0.71}^\blacktriangle$        \\ \cline{1-8}
\end{tabularx}
}
\end{table}
\begin{table}[t]
\centering
\caption{Overall results on CODEC, TREC News 2021, and TREC Core 2018. 
\textbf{nDCG and Precision measures at cut-off rank 20}.}
\label{tab:res-tab-2}
\scalebox{0.75}{
\begin{tabularx}{\textwidth}{|l|CCC|CCC|CCC|}

\hline
            & \multicolumn{3}{c|}{\textbf{CODEC}} & \multicolumn{3}{c|}{\textbf{TREC News 2021}} & \multicolumn{3}{c|}{\textbf{TREC Core 2018}} \\ \hline
& \textbf{MAP}  & \textbf{nDCG}  & \textbf{Prec} & \textbf{MAP} & \textbf{nDCG} & \textbf{Prec} & \textbf{MAP} & \textbf{nDCG} & \textbf{Prec} \\ \hline
BM25+RM3    & 0.36$^\star$    & 0.38$^\star$     & 0.43$^\star$ & 0.47$^\star$                      & 0.48$^\star$                      & 0.58$^\star$ & 0.31$^\star$                      & 0.45$^\star$                      & 0.46$^\star$ \\ \hline
RankT5 & 0.38$^\blacktriangle$    & 0.44$^\blacktriangle$     & 0.45$^\blacktriangle$ & 0.27$^\blacktriangledown$                      & 0.31$^\blacktriangledown$                      & 0.36$^\blacktriangledown$ & 0.22$^\blacktriangledown$& 0.33$^\blacktriangledown$                      & 0.33$^\blacktriangledown$ \\ 
RoBERTa     & 0.36    & 0.38     & 0.43 & 0.47                      & 0.52$^\blacktriangle$                      & 0.58 & 0.26$^\blacktriangledown$                      & 0.36$^\blacktriangledown$                      & 0.39$^\blacktriangledown$ \\ 
DeBERTa     & 0.39$^\blacktriangle$    & 0.45$^\blacktriangle$     & 0.46$^\blacktriangle$ & 0.43$^\blacktriangledown$                      & 0.47                      & 0.56 & 0.35$^\blacktriangle$                      & 0.52$^\blacktriangle$                      & 0.51$^\blacktriangle$ \\ 
ELECTRA     & 0.32$^\blacktriangledown$    & 0.32$^\blacktriangledown$     & 0.41$^\blacktriangledown$ & 0.23$^\blacktriangledown$                      & 0.46$^\blacktriangledown$                      & 0.53$^\blacktriangledown$ & 0.24$^\blacktriangledown$                      & 0.35$^\blacktriangledown$                      & 0.36$^\blacktriangledown$ \\ 
ConvBERT    & 0.38$^\blacktriangle$    & 0.44$^\blacktriangle$     & 0.46$^\blacktriangle$ & 0.44$^\blacktriangledown$                      & 0.47                      & 0.55 & 0.32                      & 0.50$^\blacktriangle$                       & 0.49$^\blacktriangle$ \\ 
KNRM        & 0.30$^\blacktriangledown$    & 0.30$^\blacktriangledown$      & 0.34$^\blacktriangledown$ & 0.14$^\blacktriangledown$                      & 0.15$^\blacktriangledown$                      & 0.19$^\blacktriangledown$ & 0.11$^\blacktriangledown$                      & 0.14$^\blacktriangledown$                      & 0.16$^\blacktriangledown$ \\ \hline
ERNIE       & 0.39$^\blacktriangle$    & 0.46$^\blacktriangle$     & 0.47$^\blacktriangle$ & 0.48                      & 0.53$^\blacktriangle$                      & 0.60$^\blacktriangle$  & 0.34$^\blacktriangle$                      & 0.52$^\blacktriangle$                     & 0.51$^\blacktriangle$ \\ 
EDRM       & 0.30$^\blacktriangledown$    & 0.29$^\blacktriangledown$     & 0.35$^\blacktriangledown$  & 0.09$^\blacktriangledown$                      & 0.10$^\blacktriangledown$                       & 0.13$^\blacktriangledown$ & 0.09$^\blacktriangledown$                      & 0.10$^\blacktriangledown$                       & 0.13$^\blacktriangledown$    \\ 
MaxSimCos   & 0.33$^\blacktriangledown$    & 0.33$^\blacktriangledown$     & 0.40$^\blacktriangledown$ &  0.20$^\blacktriangledown$                         &  0.22$^\blacktriangledown$                         & 0.30$^\blacktriangledown$      &  0.10$^\blacktriangledown$                         &  0.12$^\blacktriangledown$                         & 0.14$^\blacktriangledown$     \\ \hline
TREC Best    
& --           & --          & -- &0.43$^\blacktriangledown$	& --	& -- & 0.43$^\blacktriangle$ & -- & 0.61$^\blacktriangle$       \\ \hline
\textbf{DREQ}       & $\mathbf{0.58}^\blacktriangle$    & $\mathbf{0.50}^\blacktriangle$     & $\mathbf{0.68}^\blacktriangle$  & $\mathbf{0.76}^\blacktriangle$    & $\mathbf{0.56}^\blacktriangle$     & $\mathbf{0.84}^\blacktriangle$  & $\mathbf{0.58}^\blacktriangle$            & $\mathbf{0.66}^\blacktriangle$           & $\mathbf{0.71}^\blacktriangle$        \\ \hline
\end{tabularx}
}
\end{table}

\begin{figure}[t]
    \centering
    \begin{subfigure}{0.45\textwidth}
        \centering
        \includegraphics[width=1\linewidth]{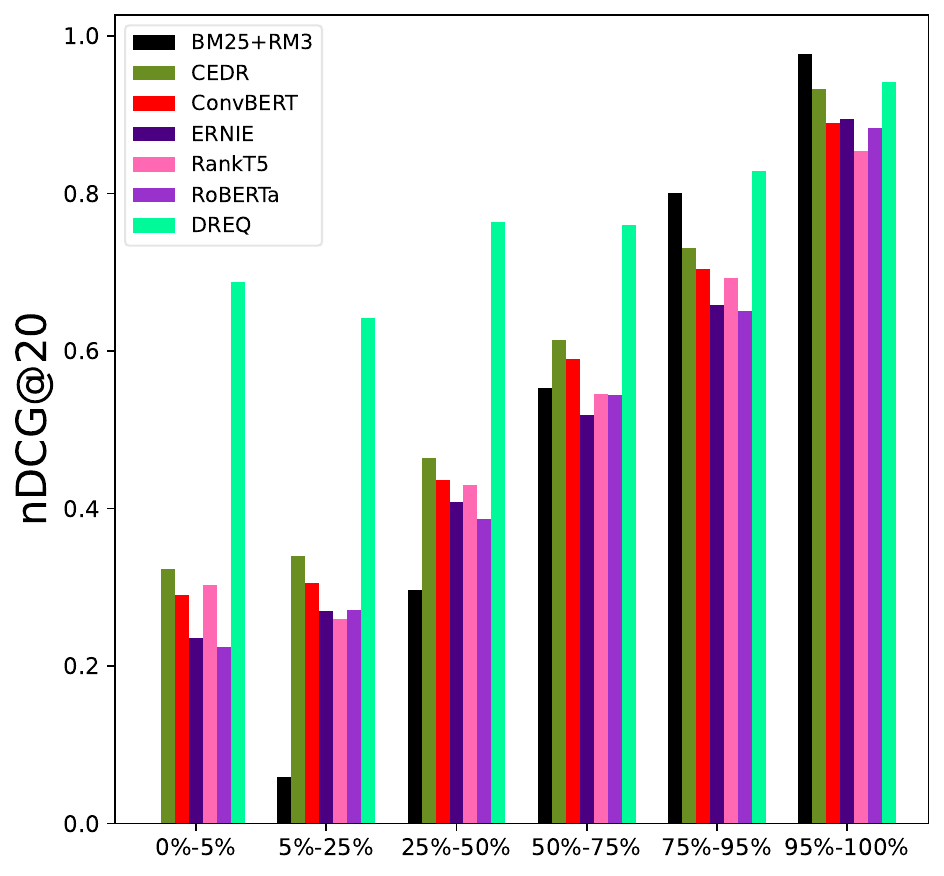} 
        \caption{Using performance of BM25+RM3}
        \label{fig:diff-test-all}
    \end{subfigure}
    \hfill
    \begin{subfigure}{0.45\textwidth}
        \centering
        \includegraphics[width=1\linewidth]{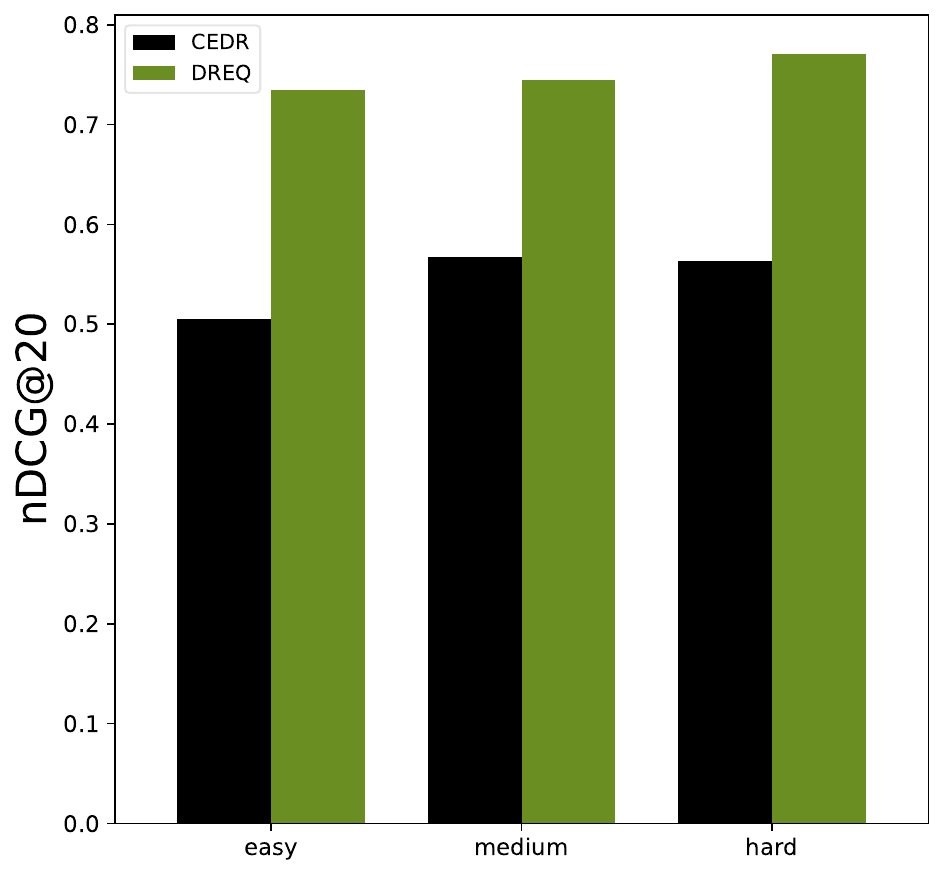}
        \caption{Using WIG QPP Method}
        \label{fig:diff-test-cedr-vs-dreq}
    \end{subfigure}
    \caption{Difficulty test for nDCG@20 on Robust04 (title). \texttt{DREQ} improves performance for the most difficult queries.
    }
    \label{fig:diff-test}
\end{figure}
In this section, we explore the following research question: \textbf{(RQ1)} \textit{How does \texttt{DREQ} perform in comparison to state-of-the-art document re-ranking methods? Which type of queries does it help the most?}
From Tables \ref{tab:res-tab-1} and \ref{tab:res-tab-2}, we observe that \texttt{DREQ} outperforms all baselines in terms of all evaluation metrics on all datasets. For example, on Robust04 (title), the best performing baseline is CEDR which obtains $\text{nDCG@20}=0.55$; however, \texttt{DREQ} improves performance by 36\% over CEDR and achieves $\text{nDCG@20}=0.75$. While CEDR improves nDCG@20 of the candidate set (BM25+RM3) by 25\% (from 0.44 to 0.55), our model \texttt{DREQ} improves this by 70\% (from 0.44 to 0.75). 

\textbf{Query-level Analysis.} We further delve into the source of performance improvements by analyzing the results at the query-level. We categorize the queries into different levels of difficulty using (a) the performance of the BM25+RM3 candidate ranking method, and (b) Weighted Information Gain (WIG) \cite{yun2007query} Query Performance Prediction (QPP) method. Results shown in Figure \ref{fig:diff-test}.

In Figure \ref{fig:diff-test-all}, we place the 5\% most difficult queries (according to nDCG@20) for BM25+RM3, the candidate ranking method, on the left, and the 5\% easiest ones on the right. 
Remarkably, \texttt{DREQ} enhances the performance of the most challenging queries (bins 0--25\%). For example, there are no relevant documents among the top-20 for queries in bin 0--5\% (nDCG@20=0.0); however, \texttt{DREQ} improves (helps) the performance of these queries, achieving an nDCG@20 of 0.70 (twice of CEDR which achieves 0.35). Specifically, we find that \texttt{DREQ} improves the performance of 210 of the 250 queries whereas CEDR only helps 160 queries. 
%

In Figure \ref{fig:diff-test-cedr-vs-dreq}, we specifically study how the reranking pipeline \texttt{BM25+RM3 >> DREQ} compares to the best performing baseline pipeline \texttt{BM25+RM3 >> CEDR}. For this, we use a well-known QPP method called WIG which provides a measure of the effectiveness of a query in retrieving relevant documents: A higher WIG score means the query is more effective (easier) while a lower score means the query is less effective (harder). Based on this, we divide the query set into three levels of difficulty: easy, medium, and hard. Once again, we observe that compared to CEDR, \texttt{DREQ} improves performance for the most challenging queries. 

\textbf{Example.} We further examine some of the challenging queries helped by \texttt{DREQ} from bin 0--5\% in Figure \ref{fig:diff-test-all}. One such query is ``Behavioral Genetics''. For this query, a highly relevant document (according to NIST assessments) 
is placed at rank 434 in the candidate set. However, we find that \texttt{DREQ} promotes this document to rank 1 (CEDR places this document at rank 212). Upon analyzing the entity ranking for this query, we find that entities directly relevant to the query, such as \textit{Sensorineural Hearing Loss}, are assigned higher scores, while non-relevant entities, such as \textit{Home for the Holidays}, receive lower scores. By utilizing these entity scores to weigh the entity embeddings, the model can emphasize the most important entities for the query, thereby enhancing the contribution of these entities to the hybrid document embedding. This, in turn, helps the model understand the nuances of relevance for the query, thereby improving the document's ranking position. 
We discuss contribution of entities in Section \ref{subsec:Contribution of Entities}.


\textbf{\textit{Take Away.}} To answer \textbf{RQ1}, \texttt{DREQ} achieves new state-of-the-art results for the document re-ranking task on four document ranking benchmarks across five diverse query sets by promoting relevant documents to the top of the ranking, thereby improving the precision at the top ranks of the candidate set. Our experiments show that \texttt{DREQ} is particularly adept at handling the most challenging queries, a key advantage over competing models. This demonstrates the robustness of \texttt{DREQ}, as it not only improves the performance of ``easy'' queries but also significantly boosts the performance of the most ``difficult'' ones. This is crucial for real-world applications where queries may be of varying difficulty. 
\subsection{Contribution of Entities in \texttt{DREQ}}
\label{subsec:Contribution of Entities}

Entities lie at the heart of our approach. Hence, in this section, we ask: \textbf{(RQ2)} \textit{What is the contribution of entities in \texttt{DREQ}? How does the performance change if we change the underlying entity ranking system?} Furthermore, \textbf{(RQ3)} \textit{How does changing the entity weighing method alter the performance of \texttt{DREQ}?}
In the following, \textbf{we discuss results with respect to only Robust04 (title)}. 

\textbf{Ablation Study.} To evaluate the contribution of the entity component to our model's performance, we first conduct an ablation study by removing the entity embeddings and retraining the model using only the other components. Our results reveal that, when entities are removed from \texttt{DREQ}, there's a marked performance drop. The nDCG@20 score plunges to 0.26, a decrease from the 0.75 score of the full model, and even below the 0.44 of the BM25+RM3 candidate set.
For a clearer context, it is worth noting that the highest nDCG@20 score achieved by the best baseline model, CEDR, is 0.55. Performing the difficulty test described above using the BM25+RM3 candidate ranking, we find that \texttt{DREQ} without entities helps 74 queries; in contrast, the full \texttt{DREQ} model helps 210.

These results underline the pivotal role of entities in \texttt{DREQ}. The marked reduction in performance, when compared to both the full model and the best baseline, indicates that entities are integral to the model's effectiveness in document ranking and relevance interpretation. The comparison with CEDR, the best performing baseline, underscores the significance of incorporating entities, as even CEDR fails to match the nDCG@20 score of the full \texttt{DREQ} model. This suggests that entities are not only essential for \texttt{DREQ} but could also be a valuable addition to other document ranking models to enhance their performance.

\textbf{Alternative Entity Ranking.} As we use the rank scores of entities to weigh the entity embeddings when learning the hybrid document embedding, we study the effect of the entity ranking system on the performance of \texttt{DREQ}. We replace our supervised BERT-based entity ranking system with 
(1) BM25, a sparse model and, (2) GEEER \cite{gerritse2020graph}, a recently proposed, state-of-the-art, entity re-ranking method. 
GEEER first computes the (Wikipedia2Vec \cite{yamada-etal-2020-wikipedia2vec}) embedding-based score for an entity $E$ in a given candidate set of entities as a weighted sum: $\text{S}_\text{emb}(E,Q) = \sum_{e \in Q} C(e) \cdot \text{cos}(\vb{E}, \vb{e})$, where $C(e)$ is the confidence score of entity $e \in Q$ obtained from an entity linker. 
The final score is obtained by interpolating this embedding-based score with the score of the entity $E$ obtained using a retrieval model. In this work, we use BM25 as the retrieval model in GEEER.
%
%

In the results, we find that incorporating the BM25 entity ranking into \texttt{DREQ} leads to a decline in performance, with the nDCG@20 dropping from 0.75 to 0.44. Similarly, 
substituting our BERT-based entity ranking system with GEEER causes a decrease in the nDCG@20 for \texttt{DREQ}, from 0.75 to 0.42.
Our results 
underscore the importance of carefully selecting an entity ranking system that is well-suited to the specific requirements of the task. It is particularly important to note that merely using an off-the-shelf entity ranking system is insufficient; it is necessary to train the system specifically to predict entities that are likely to be mentioned in documents relevant to the query.


\textbf{Alternative Entity Weighing.} The \texttt{DREQ} model intrinsically uses entity scores, expressed as probabilities, to weight the entity embeddings within documents. Acknowledging the pivotal role these probabilistic weightings have in shaping the model's performance, we perform a series of experiments to examine various alternative weighting schemes.

First, \textbf{we remove entity weightings}, treating all entities with equivalent importance within the document's representation ($w_e=1$ in Equation \ref{eq:entity-centric-document-emb}). This leads to a significant performance decrease in \texttt{DREQ}, from an nDCG@20 of 0.75 to 0.70. The difficulty test, using the BM25+RM3 candidate ranking, shows that \texttt{DREQ} using uniform weights helps 187 queries, whereas the original model helps 210. 
This underscores the pivotal role of entities in \texttt{DREQ}. Their embeddings, when appropriately weighted by relevance probabilities, significantly amplify the semantic alignment between queries and documents, attesting to the strategic decision to utilize them. Their importance becomes evident when the individual significance is disregarded, leading to a diluted representation that hampers the model's performance. Furthermore, it's clear that the ranking system's ability to discern and prioritize relevant entities is instrumental to \texttt{DREQ}'s success. 

Additionally, we explore the \textbf{reciprocal rank (RR) of entities} as an alternative weighting mechanism for entity embeddings ($w_e=\frac{1}{rank(e)}$ in Equation \ref{eq:entity-centric-document-emb}). The results are telling: nDCG@20 plummets from 0.75 to 0.54. The RR method assumes a precipitous decrease in the importance of entities based on rank. However, this isn't always fitting. Take, for instance, the query ``Black Bear Attacks''. We find that the entity \textit{George Parks Highway}, with a score of 0.99, is ranked 71st, while \textit{Cantwell, Alaska}, with a nearly identical score of 0.98, stands at 106th rank. Such negligible probability differences juxtaposed against substantial rank disparities highlight the potential limitations of the RR system. It underscores the risk of undervaluing entities that, by probability, are nearly as pertinent as top-tier entities.

\textbf{\textit{Take Away.}} To answer \textbf{RQ2}, the significant role of entities in the performance of \texttt{DREQ} is undeniable, as evident from the ablation study and alternative entity ranking experiments. 
The experiments indicate that entities are integral to the model's effectiveness, and that the entity ranking system's ability to discern and prioritize query-relevant entities is instrumental to the success of \texttt{DREQ}. However, it is essential to carefully select and train the entity ranking system to predict entities likely to be mentioned in documents relevant to the query.

In addressing \textbf{RQ3}, our experiments unveil crucial insights. The inherent design of \texttt{DREQ}, which uses entity scores as probabilities to weight the entity embeddings, stands out as a potent approach, demonstrated by superior performance metrics. When entities are uniformly weighted, stripping them of their probabilistic distinctions, the model's performance takes a tangible hit, underscoring the model's dependence on the nuanced gradation of entity relevance. The experiment with RR as weights further cements this observation. While RR attempts to rank entities based on their ordinal positions, it fails to adequately capture the subtleties in actual relevance, particularly when entities with near-identical probabilities have significantly differing ranks. The experiments not only re-emphasize the instrumental role of entities in \texttt{DREQ} (\textbf{RQ2)}, but also the imperativeness of a precise entity ranking and weighting mechanism.


\section{Conclusion}
\label{sec:Conclusion}


We introduce \texttt{DREQ}, an entity-oriented document re-ranking model that taps into the semantic richness of entities in candidate documents. While many neural IR models overlook the differential importance of entities within documents, \texttt{DREQ} prioritizes query-relevant entities, crafting a nuanced, query-specific entity-centric representation. This is then melded with the document's text-centric representation to form a \textit{hybrid} embedding, capturing \textit{query-specific insights}.
%
%
%
We provide compelling evidence of the effectiveness of \texttt{DREQ} in the document re-ranking task across multiple benchmarks and diverse query sets. \texttt{DREQ} achieves new state-of-the-art results by 
enhancing precision at the top ranks of the candidate set. Its proficiency in handling challenging queries demonstrates its robustness and underscores its potential for real-world applications where queries may vary in difficulty. We also demonstrate the pivotal role of entities, and the importance of meticulously selecting and weighing entities in \texttt{DREQ}'s performance. 

Overall, our work contributes significantly to the ongoing efforts to enhance the effectiveness of information retrieval systems. The \texttt{DREQ} model, with its innovative approach to incorporating entities and its proven ability to outperform existing methods, presents a promising avenue for further exploration and development in the field of information retrieval.

\bibliographystyle{splncs04}
\bibliography{main}
\end{document}